\documentclass{article}

\newif\ifresearchgate

\researchgatetrue    

\usepackage{hyperref}

\newcommand{\PublicationYear}{2026}
\newcommand{\PublicationVenue}{IEEE ICASSP 2026}
\newcommand{\PublicationDOI}{10.1109/ICASSP55912.2026.11463564}

\usepackage{spconf,amsmath,graphicx}
\usepackage{cite}
\usepackage{nicefrac}
\usepackage{multirow}
\usepackage{booktabs}
\usepackage{caption}

\title{ICASSP 2026 Radar Acoustic Speech Enhancement (RASE) Challenge}
\name{Zhi-Wei Tan$^{1}$, V. G. Reju$^{1}$, Ritesh C. Tewari$^{1}$, Ruotong Ding$^{1}$, Patrick A. Naylor$^{2}$, and Andy W. H. Khong$^{1,3}$
}
\address{$^{1}$School of Electrical and Electronic Engineering, Nanyang Technological University, Singapore \\
$^{2}$Dept. of Electrical and Electronic Engineering, Imperial College London, UK \\
$^{3}$Lee Kong Chian School of Medicine, Nanyang Technological University, Singapore 
}
\makeatletter
\renewcommand{\section}{\@startsection{section}{1}{\z@}%
  {-1.2ex plus -0.5ex minus -0.2ex}
  {0.5ex plus 0.2ex}
  {\normalfont\bfseries\large}}
\renewcommand{\subsection}{\@startsection{subsection}{2}{\z@}%
  {-0.6ex plus -0.4ex minus -0.2ex}%
  {0.3ex plus 0.1ex}%
  {\normalfont\bfseries\normalsize}}
\makeatother

\begin{document}


\ifresearchgate

\makeatletter
\let\PaperTitle\@title
\makeatother

\begin{titlepage}

\centering

\vspace*{2cm}

{\LARGE\bfseries
\PaperTitle
\par}

\vspace{0.8cm}

{\large\itshape
Author-Accepted Manuscript
\par}

\vspace{1.5cm}


\vspace{2cm}

\begin{minipage}{0.85\textwidth}
\raggedright

\section*{Publication Information}

This paper was published in the Proceedings of the
2026 IEEE International Conference on Acoustics,
Speech and Signal Processing (ICASSP 2026).

\vspace{0.5cm}

\section*{Suggested Citation}

\noindent
If you use this work, please cite the official
IEEE ICASSP 2026 using the DOI below.

\vspace{0.5em}

\textbf{Official Published Version:}

\url{https://doi.org/\PublicationDOI}

\vspace{1cm}

\section*{IEEE Copyright Notice}

\copyright~\PublicationYear~IEEE.

Personal use of this material is permitted.
Permission from IEEE must be obtained for all other uses,
in any current or future media, including
reprinting/republishing this material for advertising
or promotional purposes, creating new collective works,
for resale or redistribution to servers or lists,
or reuse of any copyrighted component of this work
in other works.

\vspace{1cm}

\section*{Manuscript Version}

This document contains the author-accepted manuscript
of the paper published at \PublicationVenue.

The final published version is available through
IEEE Xplore at the DOI provided above.

\end{minipage}

\end{titlepage}

\clearpage

\fi

%
\maketitle

\begin{abstract}
This paper presents the ICASSP 2026 Radar Acoustic Speech Enhancement~(RASE) Grand Challenge, the first community challenge dedicated to benchmarking speech enhancement from millimeter-wave~(mmWave) radar vibrometry. The challenge targets realistic through-glass sensing scenarios where microphones are ineffective or impractical. We introduce a curated dataset of paired radar and clean speech recordings, standardized preprocessing, a baseline system, and a secure Codabench-based evaluation framework with unified objective metrics. The RASE Challenge establishes the first public benchmark for radar-based speech enhancement and aims to foster reproducible research in this emerging field.
\end{abstract}
\begin{keywords}
Radar Vibrometry, Speech Enhancement, Speech Acquisition, ICASSP Grand Challenge
\end{keywords}
\section{Introduction}
\label{sec:intro}
Speech enhancement is an ever more important area of research~\cite{Lin2012,Zhang2025,Shetu2025,Thaleiser2025, zhiwei2025, Yang2025}. Recently, through-barrier speech capture has been proposed for scenarios where microphone systems fail or cannot be deployed effectively.
To this end, millimeter-wave~(mmWave) radar vibrometry offers a promising solution for non-contact signal acquisition, high spatial resolution, and its ability to penetrate barriers and recover speech-induced surface vibrations~\cite{Tan2025}. However, recovering high-quality, wideband speech from radar measurements remains challenging\textemdash radar-derived signals are often band-limited, suffer from low SNR, and influenced by the mechanical properties of the reflecting surface and sensing geometry~\cite{Zhao2016,Tian2013}. These factors introduce distortions that differ from those encountered in conventional speech enhancement research.

Despite growing interest in radar-based speech processing, prior work has largely relied on bespoke hardware prototypes without standardized datasets and benchmarks~\cite{WaveEar2019, Ozturk2021, RadioMic2023}, limiting reproducibility and the dissemination of research outcomes within the broader speech enhancement community. Moreover, there is no widely-adopted benchmark that incorporates realistic radar acquisitions with paired clean references and standardized evaluation protocols, making it challenging to evaluate and compare modeling approaches fairly.

The ICASSP 2026 Radar Acoustic Speech Enhancement~(RASE) Grand Challenge serves as an open platform to bridge these gaps and advance research into radar-based speech processing by establishing the first community benchmark for through-glass radar–speech enhancement. The challenge introduces a dataset of paired radar–speech recordings, baseline processing pipelines and modeling code, and a standardized evaluation framework designed to stress-test generalization in the presence of severe sensing distortions.

Specifically, this challenge provides a streamlined, publicly-available dataset along with a fully reproducible baseline implementation to promote transparent and fair benchmarking. The dataset comprises pre-processed radar-to-audio signals that enable the speech processing community to focus on developing speech enhancement models. The provided baseline and codebase establish a well-founded and well-documented reference system for participants to implement their models. Baseline architecture and training details, based on WaveVoiceNet~\cite{WaveEar2019}, are provided in the publicly available implementation and documentation,\footnote{https://github.com/RASE-Challenge/challenge\_baseline2026} serving as a reproducible reference for benchmarking. This challenge establishes a unified benchmark that supports objective comparison and fosters collective innovation in this research field.



\begin{table*}[!h]
    \centering
    \begin{tabular}{|l l|cccc|cc|c|}
        \hline
        \multirow{2}{*}{Rank}&
          \multirow{2}{*}{Team}& 
         \multicolumn{4}{c|}{Metrics}& \multicolumn{2}{c|}{Avg. Score (scenarios)} & \multirow{2}{*}{Overall}\\
        \cline{3-8}
         & & PESQ & ESTOI & DNSMOS & CS-MFCC   & (a) Direct & (b) Secondary & \\ \cline{1-9}
        1st & WHU IASP & $\mathbf{1.541}$
        & $0.248$
        & $\mathbf{2.650}$
        & $0.598$
        & $\mathbf{0.4129}$
        & $\mathbf{0.2966}$
        & $\mathbf{0.3431}$
        \\
    
        2nd & ZJU ShulinZhu & $1.478$ & $\mathbf{0.262}$ & $2.338$ & $0.624$ & $0.3917$ & $0.2895$ &  $0.3304$ \\
        3rd & USTC NERCSLIP & $1.318$ & $0.194$ & $1.673$ & $0.625$ & $0.3176$ & $0.2236$& $0.2612$ \\
        4th & Team Quantz & $1.306$ & $0.183$ & $1.580$ & $\mathbf{0.627}$ &$0.3086$ & $0.2152 $ & $0.2526$ \\
        5th & Origin Research & $1.301$ & $0.183$ & $1.642$ & $0.608$ & $0.3025$ & $0.2183$ & $0.2520$ \\
        6th & SSSL  & $1.317$ & $0.157$ & $2.040$ & $0.518$ & $0.3075$ & $0.2074$ & $0.2474$ \\
        7th & Augm. Listen. Lab & $1.209$ & $0.163$ & $2.121$ & $0.506$ & $0.3224$ & $0.1853$ & $0.2401$ \\
        \addlinespace[-0.4pt]
        \midrule
        \addlinespace[0pt]
        N/A & Baseline & $1.279$ & $0.166$ & $1.436$ & $0.604$ & $0.2820$& $0.1994$& $0.2324$ \\
        \addlinespace[0pt]
        \midrule
        \addlinespace[0pt]
        8th & SRIB Language AI & $1.162$ & $0.026$ & $1.449$ & $0.449$ & $0.1698$ & $0.1477$ & $0.1565$ \\ 
        
        \hline

    \end{tabular}
    \caption{Objective results of the top eight models submitted to the RASE 2026 Challenge. Higher values indicate better performance for all metrics. Team affiliations:  
WHU IASP~(Wuhan University),  
ZJU ShulinZhu~(Zhejiang University),  
USTC NERCSLIP~(University of Science and Technology of China),  
Team Quantz~(Indian Institute of Science),  
Origin Research~(Origin Research),  
SSSL~(Korea Advanced Institute of Science and Technology),  Augmented Listening Lab~(Stony Brook University), and SRIB Language AI~(Samsung R\&D Institute Bangalore). Average scores for~(a)~direct diaphragm and~(b)~secondary-surface
vibration are reported, with the weighted overall score in the last column.}
    \label{tab:placeholder}
\end{table*}
\section{Challenge Description and Evaluation} 
\label{sec:ChallengeDescription}

Unlike existing within-the-same-closure speech enhancement approaches\cite{Khong2008, Tao2025}, the RASE Challenge evaluates speech enhancement methods for through-glass radar-based sensing. Here, our goal is to develop models that reconstruct intelligible, full-bandwidth speech from mmWave FMCW radar measurements that capture only speech-induced surface vibrations. The task reflects realistic deployment scenarios in which the radar is placed outside a glass-walled room and acquires vibration amplitude via radar vibrometry signals.


The dataset comprises
speech utterances played via a loudspeaker through a glass panel under two scenarios:
(a) direct diaphragm vibration, where the radar measures vibrations from the loudspeaker diaphragm; and (b) secondary-surface vibration, where a thin aluminium foil placed near the loudspeaker vibrates indirectly, generating a weaker signal. The data acqusition system was designed to operate across both sensing conditions to generate enhanced speech suitable for human listening and downstream tasks.

Both scenarios were recorded using a 77~GHz mmWave FMCW radar~(TI~AWR2243BOOST) with a chirp rate of~8~kHz. 
In particular, for FMCW radar vibrometry, speech is acquired by sensing vibration-induced micro-displacements of a target surface encoded in the radar instantaneous-frequency~(IF) signal. Received IF radar signals were then mapped into vibration-derived speech signals by estimating the IF phase variation~\cite{basak2022mmspy, guo2021measuring}. The participants were then provided the vibration-derived speech signals and the time-aligned clean speech signals, both in standard audio playback~.wav format at 8~kHz sampling frequency. 

The training and validation utterances were sourced from the LibriSpeech corpus~\cite{panayotov2015librispeech}, ensuring sufficient linguistic and acoustic diversity for model development. In contrast, the held-out test set is constructed from speech drawn from an unannounced corpus. 
This cross-corpus evaluation was intentionally designed to assess model generalizability beyond the training distribution, preventing overfitting to LibriSpeech-specific characteristics to reflect robustness.



The rules of this challenge were  designed according to the principles of: fairness, transparency, and focusing on the development of speech enhancement models. To ensure fair comparison, all teams were required to train their models from scratch~(i.e., no pre-trained weights) using only the provided training and development data; the use of external data or augmentation is prohibited to avoid unfair advantages from access to large corpora or pre-trained models. To prevent overfitting to the test data, each team is limited to a maximum of three test-phase submissions.

The held-out test data were not released and evaluation was conducted through the Codabench platform~\cite{Codabench} with a secure server-side execution ensuring standardized and tamper-resistant benchmarking. To support model development, participants were granted access to an unrestricted development-trial portal and a validation phase with up to thirty submissions for code verification before the test phase. Templates and a single-line execution command further facilitate reliable packaging and submission.




Submitted models were evaluated on a held-out test set comprising both sensing conditions: direct diaphragm vibration and secondary-surface vibration. For each utterance, we compute four complementary objective measures capturing intelligibility, perceptual quality, and spectral consistency. The challenge score is defined as {\small
\begingroup
\setlength{\abovedisplayskip}{5pt}
\setlength{\belowdisplayskip}{5pt}
\setlength{\abovedisplayshortskip}{5pt}
\setlength{\belowdisplayshortskip}{5pt}
\[
\text{Avg. Score}=\frac{\text{PESQ}_n + \text{ESTOI} + \text{CS-MFCC} + \text{DNSMOS}_n }{4},
\]
\endgroup
}where $\text{PESQ}_n\in[0,1]$ and $\text{DNSMOS}_n\in[0,1]$ denote linearly normalized PESQ~\cite{Rix2001} and DNSMOS~\cite{Reddy2020DNS}, respectively. The extended short-time objective intelligibility~(ESTOI)~\cite{ESTOI} measures intelligibility and~$\text{CS-MFCC}\in[-1,1]$ quantifies spectral and phonetic consistency via cosine similarity between Mel-frequency cepstral coefficient~(MFCC). A single ranking criterion is obtained by combining metrics using a~40\% weighting for the simpler direct-diaphragm scenario and~60\% for the secondary-surface scenario. 

 

\section{Results and Conclusion}
\label{sec:results}
Table~\ref{tab:placeholder} summarises the performance of the submitted models evaluated on the held-out test set. Based on the overall score, Team WHU IASP, ZJU ShulinZhu, USTC NERCSLIP, Quantz, and Origin Research emerge as the top five teams, with SSSL and Augm. List. Lab. ranked 6th and 7th, respectively. These teams outperformed the WavVoiceNet baseline~\cite{WaveEar2019}, demonstrating their effectiveness.
The RASE 2026 Challenge represents, to our knowledge, the first community challenge for radar-acquired speech enhancement, with the results establishing the first public benchmark in this area. The accompanying database, baseline system, and evaluation framework are released via our website\footnote{https://rase-challenge.github.io/RASE2026-Challenge/} and Codabench portal\footnote{https://www.codabench.org/competitions/10857/}, enabling standardized and reproducible benchmarking.
We thank all participants for their efforts and submissions.
\newpage

\bibliographystyle{ieeetr}
\bibliography{library}

\end{document}